\documentstyle {article}

\begin{document}

{\bf About Perpetuum Mobile without Emotions.}

\

{\it A.V.Nikulov,}

Institute of Microelectronics Technology and High Purity Materials, Russian
Academy of Sciences, 142432 Chernogolovka, Moscow District, RUSSIA

\

One of the oldest science problems - possibility of the perpetuum mobile is discussed. The interest to this problem was provoke a result, published recently, which contradicts to the second law of thermodynamics. According to this result, the thermal fluctuations can induce a voltage with direct component in a inhomogeneous superconducting ring at an unaltered temperature corresponded to the resistive transition of the ring segment with the lowest critical temperature. This result arises from obvious statements: 1) the switching of a ring segment $l_{b}$ into and out of the normal state, while the rest of the ring (segment $l_{a}$) remains superconducting, can induced a voltage with dc component (It is shown that, in spite of the wide spread opinion, this statement is correct because the superconductivity is a macroscopic quantum phenomena); 2) the thermal fluctuations switch the mesoscopic ring segment $l_{b}$ with lowest critical temperature $T_{sb}$ into and out of the normal state at $T \simeq T_{sb}$, while the rest of the ring remains superconducting if $T_{sa} > T \simeq T_{sb}$. In order to resolve the contradiction between these obvious statements and the second law of thermodynamics a possibility of the second order perpetuum mobile is considered theoretically. It is shown that from two type of the perpetuum mobile, only type "b" and only in quantum systems is possible. According to the presented interpretation, the total entropy, as the measure of the chaos, may be systematically reduced in some quantum system because a "switching" between the classical and quantum mechanics is possible. Instruction for the making of the perpetuum mobile is enclosed.

\

\section{Introduction}

Physics is not lyricism. Any sentiment is inappropriate here. Nevertheless some physical problems provoke emotion. The first of these problems is the perpetuum mobile. Any statement on a possibility of perpetuum mobile provokes first of all the sense of distrust. According to the dominant opinion this problem is once and for all decided. Almost all scientists, during more than two centuries, are fully confident in the impossibility of any perpetuum mobile. I, as well as other grave scientists, was sure that only madman may be in earnest about a possibility of any perpetuum mobile. But a result, which I have obtained recently, has compelled me to change my point of view. According to this result the chaotic energy of thermal fluctuation can be transformed to the electric energy of a direct current at an unaltered temperature by means of a mesoscopic inhomogeneous superconducting ring.

This result arises from obvious statements: 1) the switching of a ring segment $l_{b}$ into and out of the normal state, while the rest of the ring (segment $l_{a}$) remains superconducting, can induced a voltage with dc component; 2) the thermal fluctuations switch the mesoscopic ring segment $l_{b}$ with lowest critical temperature $T_{sb}$ into and out of the normal state at $T \simeq T_{sb}$, while the rest of the ring remains superconducting if $T_{sa} > T \simeq T_{sb}$. Both these statements are agreed with our modern knowledge. But it follows directly from this statements that the thermal fluctuation can induce a voltage with dc component
at an unaltered temperature corresponded to the resistive
transition of the ring segment with the lowest critical temperature. It is obvious that the dc voltage may be used for an useful work. This means that the useful work can be obtained from the chaotic energy of thermal fluctuation (i.e. from the heat energy) at unaltered
temperature (i.e. in the equilibrium state). This possibility contradicts directly to the second law of thermodynamic. And it is well known that a violation of the second law of thermodynamics means a possibility of the second order perpetuum mobile.

In order to resolve this contradiction between the obvious statements and the second law of thermodynamics I have investigated the reason of the firm belief in the impossibility of perpetuum mobile. As a result I conclude that this belief does not have a theoretical substantiation. Only argument against the perpetuum mobile is numerous unsuccessful attempts to invent it. But it is not strict argument. What could not be made yesterday can be made today. For example, the mesoscopic superconducting ring can not be made twenty years ago but it can be made at present. Now I am sure that the statements 1) and 2) are correct. The second order perpetuum mobile is possible because the chaos may be systematically reduced in some quantum system. Therefore I have published in \cite{jltp98,yalta98} the result which contradicts to the second law of thermodynamics \cite{lt22}.

But it is not enough to publish such result. No one will straight off believe that such result can be correct. Therefore I think that I ought expound in detail my arguments. I make this in the present article. In the beginning I remind briefly the history of the considered problem. After that a brief theoretical consideration of the perpetuum mobile problem is presented. In the section 4, the quantum force is introduced in order to explain the dc voltage appearance in the superconducting ring segment. In the section 5, I explain when and why the total entropy may be systematically reduced. The last section is directions for who want to make the perpetuum mobile.

I must write that the theoretical result published in \cite{jltp98} was provoked by an experimental result. But this experimental result is not published. Therefore the result \cite{jltp98} ought be considered as the theoretical prediction but not as an explanation.

\section{A few history}

The perpetuum mobile is one of the oldest problems of science. This problem
is more old than almost all foundations of modern physics. Many persons
attempted to invent a perpetuum mobile during many centuries. Such attempts
were known beginning with 13 century. The principle of the impossibility of the perpetuum mobile was postulated first by Stevin (1548-1620 years). The Paris Academy of Sciences has decided in 1775 year to do not consider any project of a perpetuum mobile. This verdict did
not have any scientific basis. The first and second laws of thermodynamics were formulated only in the next century. Nevertheless, beginning with that time, almost all (with the exception of few \cite{Berger}) scientists are sure that a perpetuum mobile is impossible.

Sometimes one says that the impossibility of a perpetuum mobile is based on the first and second laws of thermodynamics. But it is not quite right. It is more right to say that the second law of thermodynamics is based on the statement on the impossibility of a perpetuum mobile.
The first formula of this law - Carnot's principle - was proposed in 1824 year. Carnot wrote that the useful work can not be obtained from the heat energy at unaltered temperature (in the equilibrium state) because in the opposite case the perpetuum mobile is possible. Following the Carnot's idea Rudolf Clausius in 1850 year and William Thomson (Lord Kelvin) in 1851 year have proposed the formulas which are more known
now. According to the Clausius's formula the heat energy can not be
transferred from a cold body to a hot body without an expense of an
additional energy. According to the Thomson's formula it is impossible to
obtain a power-driven energy (useful work) by means of a cooling of a body with lowest
temperature. The second law of thermodynamics is formulated also as the law of entropy increase. The entropy S was introduced first in 1865 year by Rudolf Clausius as a
value which changes on $\Delta S = Q/T$ in a reversible process when a
thermodynamic body obtains the heat energy $Q$ ($Q > 0$) or gives the heat
energy ($Q < 0$). T is the Kelvin's temperature of the body. Ostwald was formulated in 1877 year the second law of thermodynamics as the impossibility of the second order perpetuum mobile. This formulas are equivalent and are used for the present. But the modern interpretation of the second law of thermodynamics, as well as the entropy, differ in essence from the old one dominated in the last century.

According to the both interpretation the entropy value, $S$, can not decrease in
a closed thermodynamic system. It does not change in the reversible process and increases in irreversible process. Any thermodynamic system tends to the equilibrium state
corresponding to a greatest $S$ value. But there is the difference between old and modern definition of the entropy which causes some contradiction between old and modern interpretation of the second law of thermodynamics.

According to the Clausius's definition, the temperatures of parts of any thermodynamic system can not differ in the equilibrium state. If the temperatures differ, $T_{1} > T_{2}$, the entropy
increases $\Delta S = Q/T_{2} - Q/T_{1}$ when the heat energy $Q$
is transferred from a hot (with $T = T_{1}$) to a cold (with $T =
T_{2}$) part of the system. Consequently, the state with $T_{1} > T_{2}$ is not equilibrium. According to the old interpretation the heat energy flows only from a hot to a cold part and any dynamic process (any transfer of the heat energy) does not take place in the equilibrium state.

Following to L.Boltzmann, J.W.Gibbs and others
we define now the entropy by the relation $S = k_{B}\ln P$. Here $k_{B}$
is the Boltzmann constant; $P$ is the statistical weight proportional to a
number of microscopic states and to the probability of macroscopic state. The maximum entropy corresponds to the maximum probability: $P = \exp(S/k_{B})$. It is easy to show  (see \cite{kittel}) that the number of microscopic states $P_{1}$ at $T_{1} = T_{2}$ is higher than $P_{2}$ at $T_{1} > T_{2}$ (at the same total internal energy). We could make the conclusion from this statement that the modern definition of the entropy is agreed with the Clausius's definition. But on other hand, $P_{1} + P_{2} > P_{1}$. The thermodynamic system can not be at the same time in the different states, with $T_{1} = T_{2}$ and $T_{1} > T_{2}$. But it can goes between these states. This process is well known as the thermal fluctuation. The heat energy is transferred from a part to other part of any thermodynamic system at $T > 0$. The $Q$ transfer from a cold $T_{2}$ to hot $T_{1}$ part of a system is hardly probable if $-\Delta S = -(Q/T_{1} - Q/T_{2}) \gg k_{B}$. But the probability $\propto \exp(Q/k_{B}T_{1} - Q/k_{B}T_{2})$ of this transfer in opposite direction does not differ strongly if $Q|T_{2}-T_{1}|/k_{B}T_{1}T_{2} < 1$ (i.e. the Clausius's formula of the second law of thermodynamics is correct to an accuracy of $\Delta S = k_{B}$). The fluctuations contradict to the old interpretation. Therefor they were interpreted as the violation of the second law of thermodynamics in the beginning of our century \cite{Smolucho}. Evidence of the fluctuation existence was one of the reasons why the modern interpretation has won the old interpretation. According to the modern interpretation, the second law of thermodynamics has probabilistic but not reliable nature. We know now that dynamic processes take place in the equilibrium state at $T > 0$. But these processes are chaotic. According to the modern interpretation, the second law of thermodynamics is the law of chaos increase. And the entropy is interpreted now as a measure of the chaos.

The second law of thermodynamics was formulated in order to describe the transformation of a heat energy $Q$ to an useful work $A$ in the heat engine. Scientists understood that the mechanical (power-driven) energy is dissipated completely in the heat at any work in a consequence of the friction, $E_{mech} \rightarrow A \rightarrow Q$. Proceeding from the impossibility of the perpetuum mobile Carnot has shown in 1824 year that the heat energy can not be transformed completely to the mechanical energy. A heater with $T_{heat}$ and a cooler with $T_{cool} < T_{heat}$ should be in any heat engine, according to Carnot. A work body obtains the heat energy $Q_{1}$ from the heater at $T = T_{max} \leq T_{heat}$ and gives the heat energy to the cooler at $T = T_{min} \geq T_{cool}$. According to the Carnot's law, the efficiency $Ef = A/Q$ of any heat engine can not exceed $Ef_{max} = 1 - T_{min}/T_{max}$. This law is considered now as the consequence of the first law of thermodynamics $A = Q_{1} - Q_{2}$ and of the second law of thermodynamics $\Delta S = Q_{2}/T_{min} - Q_{1}/T_{max} \geq 0$. The $E_{max}$ is realized in a reversible regime when $\Delta S = 0$. But these laws of thermodynamics were formulated later than the Carnot's law.

According to the Carnot's law, $Ef_{max} = 0$ in the equilibrium state and the total heat energy is systematically increased at the work because the mechanical energy dissipated in the heat energy can not be restored completely. Therefore we must use a fuel in order to obtained the useful work. The Carnot's law remains without change to our time although it's substantiation and interpretation were changed qualitatively. According to the old interpretation the work can not be obtained at $T_{max} = T_{min}$ because any dynamic process is absent in the equilibrium state. According to the modern knowledge the heat is the chaotic motion. The chaotic dynamic processes take place in any thermodynamic system at $T > 0$. The Carnot's law is connected now with the law of chaos increase. The total entropy, as the measure of the chaos, may be systematically reduced if the heat engine is possible in which $Ef = A/Q_{1} > Ef_{max} = 1 - T_{min}/T_{max}$. The reduction of the entropy at the transformation of the heat energy to the mechanical energy (or other form of ordered energy $E_{ord}$), $Ef Q_{1} \rightarrow A = \int dX F_{ord} \rightarrow E_{ord}$, is not completely compensated by the transfer of the heat energy $Q_{2}$ from the heater to the cooler if $Ef = A/Q_{1} > Ef_{max}$. ($E_{ord}$ is, for example, the kinetic energy of a flywheel or the magnetic energy $LI^{2}/2$ in a solenoid.) The chaos increase, taken place at the dissipation of the ordered energy $E_{ord} \rightarrow A = \int dX F_{dis} \rightarrow Q_{dis}$, may be completely compensated if $Ef > Ef_{max}$. (The dissipative force $F_{dis}$ is, for example, the friction force retarding the flywheel or the force decreasing the current value $I$ in the solenoid when it's resistance is not equal zero.) Consequently, the heat engine with $Ef > Ef_{max}$ is the second order perpetuum mobile: the useful work could be obtained anyhow long time without any fuel. According to the opinion dominated now, it is impossible because the total chaos can not be reduced. This
 belief, as well as the second law of thermodynamics in the old interpretation, is founded on the postulate of the perpetuum mobile impossibility. This postulate has long and rich history and, may be, therefore does not have a theoretical substantiation. This problem ought be considered theoretically at last.

\section{Theory of Perpetuum Mobile}

A possibility of a perpetuum mobile means that the useful work can be obtained anyhow long time $T$. At the useful work, as well as at any other work, an energy is transferred from a part to other part of the system. The work $A$ is the product of a force $F$ into a distance $dX$, $A = \int dX F$. Because $dX = vdt$, $A = \int_{0}^{T} dt Fv  = T<Fv>$. Here $<Fv> = \int_{0}^{T} Fv dt/T$ is the average by the time T of the product of the force $F$ into the velocity $v$. Consequently, a perpetuum mobile is possible if a process exists in which the average $<Fv>$ by anyhow long time T is not equal zero. If $F$ is the total force of a closed system and $<Fv> \neq 0$ then the first order perpetuum mobile is possible. $<Fv> \neq 0$ contradicts to the law of energy conservation (the first law of thermodynamics). I can not doubt the this law. Therefore I, as well all other scientist, am fully confident in the impossibility of the first order perpetuum mobile. Let turn to the second order perpetuum mobile.

Both the conventional heat engine and the second order perpetuum mobile do not create a new energy. They put in order the chaotic heat energy. Because the ordered energy is dissipated at any real work $E_{ord} \rightarrow T<F_{dis}v> \rightarrow Q_{dis}$ the heat energy should be transformed in the ordered energy $Q_{dis} \rightarrow T<F_{ord}v> \rightarrow E_{ord}$ in order the work can take place any long time.

$<F_{ord}v> \neq 0$ in two cases: a)if $<F_{ord}v> \neq <F_{ord}><v>$, or b) if both $<F_{ord}> \neq 0$ and $<v> \neq 0$. Thus, two type of both heat engine and second order perpetuum mobile may be: type "a", when $<F_{ord}v> \neq <F_{ord}><v>$ and type "b", when $<F_{ord}v> = <F_{ord}><v>$ but both $<F_{ord}> \neq 0$ and $<v> \neq 0$. The case a) takes place if the force $F$ and the velocity $v$ are correlated. This correlation takes place in a conventional heat engine. For example, the pressure in a steam-engine has  different value when a piston is moved in opposite directions. Therefore $<F_{ord}v> \neq 0$ although $<v> = 0$ because $<F_{ord}v> \neq <F_{ord}><v>$. But in order to achieve this correlation an controlled heat flow is used in any conventional heat engine. Such flow is possible only in the inequilibrium state. Therefore the total entropy increases both at the ordered process and at the dissipation process. This process can not be anyhow long (infinite) time because the state of thermodynamic system is changed: the total entropy increases. The heat energy $Q_{dis}$ can not be transformed completely in the ordered energy. Therefore any conventional heat engine is not the perpetuum mobile. The total state of the thermodynamic system should not change during the work of the second order perpetuum mobile. Therefore it should work in the equilibrium state because only in this state the total entropy value does not increase in time.

$F \neq 0$ in the equilibrium state only because of the fluctuation. The fluctuation is chaotic. In a chaotic process $<Fv> = m<vdv/dt> = m<dv/dt><v> = <F><v>$. There is a mathematical problem to prove that $<vdv/dt> - <v><dv/dt> = 0$ if the function $v(t)$ is chaotic. This proof is evidence of the impossibility of the type "a" perpetuum mobile.

Our last hope to invent the perpetuum mobile is the case "b". It is obvious that $<v> = 0$ in any classical (no quantum) system where all states are permitted. (There is used the reference system in which the total momentum of the considered thermodynamic system is equal zero). The probability of a state $P$ is proportional to $\exp-(E/k_{B}T)$. The energy $E$ of a state is function of $v^{2}$ in a consequence of the space symmetry. Therefore the probability $P(v) = P(-v)$ and  $<v> = \sum P(v)v + P(-v)(-v) = 0$ if all states are permitted. This argument may be considered as a theoretical substantiation of the verdict made by the Paris Academy of Sciences. The quantum mechanics was not known in 1775 year.

But it can not be considered as the evidence of the impossibility of the type "b" perpetuum mobile in our time because no all states are permitted according to the quantum mechanics. Therefore $<v> \neq 0$ in some quantum systems. One of such systems is the mesoscopic superconducting ring considered in \cite{jltp98,yalta98}. As a
consequence of the relation (see \cite{tink75})

$$v_{s} = \frac{1}{2m}(\hbar \frac{d\phi}{dr} - \frac{2e}{c}A) =
\frac{2e}{mc}(\frac{\Phi_{0}}{2\pi }\frac{d\phi}{dr} - A) \eqno{(1)}$$
the velocity of the superconducting electrons $v_{s}$ along the
circumference of a completely superconducting ring must have fixed values

$$\int_{l} dl v_{s} = \frac{e}{mc}(\Phi_{0}n - \Phi) \eqno{(2)}$$
dependent on the magnetic flux because $n = \int_{l} dl(1/2\pi )d\phi/dr$
must be an integer number since the wave function $\Psi = |\Psi |
\exp(i\phi)$ must be a simple function. Here $\Phi_{0} = \pi \hbar c/e$
is the flux quantum; A is the vector potential; m is the electron mass and
e is the electron charge; $l=2\pi r$ is the ring circumference; $r$ is the
ring radius; $\Phi = \int_{l}dlA$ is the magnetic flux contained within the
ring.

At $\Phi/\Phi_{0} \neq n$ and $\Phi/\Phi_{0} \neq n+0.5$ the permitted
states with the opposite directed velocity have different values of the
kinetic energy $E_{kin} = mv_{s}^{2}/2$. For example at $\Phi/\Phi_{0} =
1/4$ the lowest permitted velocities in a homogeneous superconducting ring
are equal $v_{s} = -\hbar/mR4$ at $n=0$ and $v_{s} = 3\hbar/mR4$ at $n=1$. The
kinetic energy of these states differ in 9 times. Therefore the
thermodynamic average of the velocity $<v_{s}>$ is not equal zero.
It is important that the motion of the superconducting condensate is circular. It is obvious the type "b" perpetuum mobile is possible only at a circular motion. Only in this case the position of the work body (superconducting condensate in the case of the considered ring) does not change without limit during anyhow long time at $<v> \neq 0$.

Thus, the result published in \cite{jltp98,yalta98} may be correct if an ordered force $F_{ord}$ exists the average value of which $<F_{ord}>$ is not equal zero in the equilibrium state. Such force exists. It acts at the closing of the superconducting state in the ring and is connected with the quantization of the generalized momentum of superconducting electrons along the ring circumference. Therefore I will call it as quantum force.

\section{A quantum force}

According to the statement 1) (see Introduction) the switching of a ring segment $l_{b}$ into and out of the normal state, while the rest of the ring (segment $l_{a}$) remains superconducting, can induced a voltage with dc component $<V_{b}> \neq 0$. $ l_{a} + l_{b} = l = 2\pi r$. This statement means that the dc voltage exists both on the switched segment $<E_{b}>
= <V_{b}>/l_{b}$ and on the superconducting segment $<E_{a}> =
<V_{a}>/l_{a}$. Because $<V_{a}> + <V_{b}> = \int_{l}dl <E> = \int_{l}dl
<-\bigtriangledown V - (1/c)dA/dt> = - (1/c)<d\Phi/dt> = 0$, $<V_{a}> = - <V_{b}>$.
Here $<E> = (\int_{0}^{T}dt E)/T$ is the average value over a long time
$T$. The value of the magnetic
flux contained within the ring $\Phi = \int_{l} dl A$ may change iteratively at the
iterative switching of the segment $l_{b}$, but $<d\Phi/dt> = 0$ when the magnetic flux $BS$ induced by an external magnet is not changed. $B$ is the magnetic induction induced by an external magnet; $S = \pi r^{2}$ is the ring area. $<E_{a}>
\neq 0$ contradicts to the wide spread opinion that the direct voltage can
not exist in any superconducting region.

This opinion is correct if only force of the electric field $F_{e} = eE$
acts on superconducting electrons. Then, according to the classical mechanics,
the velocity $v$ should increases without limit ($mv = \int_{0}^{T} dt mdv/dt = \int_{0}^{T} dt eE = e<E>T$) if $<E> \neq 0$. But if we proposed that only $F_{e}$ acts on superconducting
electrons we should conclude that any quantization is not possible. It should
be the case of the state with infinite conductivity where the average value of the generalized electron momentum along the ring circumference $<p>_{l} = l^{-1}\int_{l}dl p = l^{-1}\int_{l}dl (mv + (e/c)A) = m<v>_{l} + (e/c)\Phi/l$ can not
change because $\int_{l}dl mdv/dt = \int_{l}dl eE = \int_{l}dl e(-\bigtriangledown V -
(1/c)dA/dt) = -(e/c)d\Phi/dt$. Here $<v>_{l} = l^{-1}\int_{l}dl v$ is the average velocity along the ring circumference. But it is well known \cite{tink75} that
superconducting state differs from the state with infinite conductivity
because the superconductivity is a macroscopic quantum phenomena. The $<p>_{l}$ value can change at the transition to the superconducting state. Electrons are accelerated against the force of the electric field in this case. This takes place, for example, at the Meissner effect, the Little-Parks effect and other quantization phenomenon.

In order to describe these phenomenon using the language of the classical mechanics a quantum force $F_{q}$ should be introduced, $\int dt F_{q} = <\Delta p>_{l}$. $<p>_{l} = (e/c)\Phi/l = (e/c)BS/l$ when the ring in the normal state, because $<v>_{l} = 0$. The $<p>_{l}$ value does not change when a ring segment $l_{a}$ is switched into the superconducting state and other segment $l_{b}$ is remained in the normal state, $n_{sa} > 0$ and  $n_{sb} = 0$. $n_{sa}$ and $n_{sb}$ are the densities of superconducting electrons in the segments $l_{a}$ and $l_{b}$. This corresponds to the laws of the classical mechanics and does not require of the quantum force. Contradiction with the classical mechanics appears at the closing of the superconducting state in the ring (when $n_{sa} > 0$ and  $n_{sb} > 0$) because the momentum $2p = \hbar \bigtriangledown \phi = 2mv_{s} + \frac{2e}{c}A$ (see (1)) and the velocity of superconducting pair (see (2)) are quantized if the superconducting state is closed.
The average momentum $<2p>_{l} = l^{-1}\int_{l}dl \hbar \bigtriangledown \phi = \hbar 2\pi n/l = (2e/c)(\Phi_{0}/l)n$ and the average velocity along the ring circumference
$<v_{s}>_{l} = l^{-1} \int_{l} dl v_{s} = \frac{e}{mc}\frac{\Phi_{0}n - \Phi}{l}$ can have only permitted values when $n_{sa} > 0$ and  $n_{sb} > 0$. Therefore the average momentum should be change on $<\Delta 2p>_{l} = (2e/c)(n\Phi_{0} - BS)/l$ at the closing of the superconducting state if the magnetic flux
within the ring is not divisible by the flux quantum, $\Phi  = BS \neq n\Phi_{0}$). Consequently, the quantum force $\int_{t_{cl}} dt F_{q} = (2e/c)(n\Phi_{0} - BS)/l$ acts at the closing of the superconducting state. Here $t_{cl}$ is a time of the superconducting state closing.

The superconducting electrons in the segment $l_{a}$ are accelerated against the force of the Faraday electric field $\int_{l}dl F_{e} = e\int_{l}dl E = -ed\Phi/dt =  -eLdI/dt$. The velocity is changed from $v_{sa} = 0$ to the value determined in the stationary state, when  the current in the ring $I = I_{s}=I_{sa}=s_{a}j_{sa}=s_{a}en_{sa}v_{sa}=I_{sb}=s_{b}j_{sb}=
s_{b}en_{sb}v_{sb}$, by the relation

$$v_{sa} = \frac{e}{mc}\frac{n_{sb}}{(l_{a}n_{sb}+l_{b}n_{sa})}
(\Phi_{o}n-\Phi) \eqno{(3a)}$$
(at $s_{a} = s_{b} = s$). $s$ is the area of the wall section of the ring.
The magnetic flux inside the ring changes from $\Phi = BS$ to $\Phi = BS - LI$. In the stationary state

$$I = I_{s} = \frac{se^{2}}{mc}
\frac{n_{sa} n_{sb}}{(l_{a}n_{sb}+l_{b}n_{sa})} (\Phi_{0}n-\Phi) \eqno{(3b)}$$

The momentum $<2p>_{l}$ returns to the initial value during the decay time $L/R_{nb}$ after the transition of the segment $l_{b}$ in the normal state. Here $R_{nb}= \rho_{bn}l_{b}/s$ is the resistance of the segment $l_{b}$ in the normal state. This process corresponds to the laws of the classical mechanics. The velocity $<v_{e}>_{l}$ in the segment $l_{b}$ is decreased because of the dissipation. Therefore a charge $q = \int dt (I_{a} - I_{b})$ on the boundaries of the segments $l_{a}$ and $l_{b}$ and, as a consequence, the potential difference $V_{a}$ and $V_{b}$ appear. The electric field $E_{a} = -\bigtriangledown V_{a} - (1/c)dA/dt$ retards the superconducting electrons in the segment $l_{a}$ and $E_{b} = -\bigtriangledown V_{b} - (1/c)dA/dt$ counteracts of the dissipated force $F_{dis} = - eE_{b} + mdv_{e}/dt$ in the segment $l_{b}$. The voltage $E_{a}$ has the same direction after the transition of $l_{b}$ both in the superconducting and normal states: the superconducting electrons are accelerated against $E_{a}$ after the transition to the superconducting state and are retarded by $E_{a}$ after the transition to the normal state. Consequently $<V_{a}> \neq 0$. Thus, in spite of the wide spread opinion, the direct voltage can exist on the superconducting segments of the ring.

The quantum force acts both in inhomogeneous and in homogeneous rings. At the transition to the superconducting state of a homogeneous ring the electrons, which become superconducting, accelerate $\int dt (dv_{s}/dt) = (\hbar/2nr)(n - \Phi/\Phi_{0})$ \cite{tink75} against the electric field $E = -(L/l)dI_{s}/dt$. The work done by the quantum force $\int dx F_{q} = \int dt v_{s}F_{q}$ increases the kinetic energy $\int dt lsn_{s}m(dv_{s}/dt) v_{s} = lsn_{s}mv_{s}^{2}/2$ and the energy of magnetic field $F_{L} = \int dt (-lsn_{s}eEv_{s}) = \int dt L(dI_{s}/dt)I_{s} = LI_{s}^{2}/2$. This means that the energy of the superconducting state increases if $\Phi \neq n\Phi_{0}$. The Little-Parks effect \cite{little} is experimental evidence of this. According to the Tinkham's explanation \cite{tinkham} of this effect, the critical temperature of a superconducting tube with narrow wall depends in a periodic way on the magnetic flux value within the tube

$$T_{c}(\Phi) = T_{c}[1 - (\xi(0)/r)^{2}(n-\Phi /\Phi_{0})^{2}]
\eqno{(4)} $$
because the $|n-\Phi /\Phi_{0}|$ tends towards a minimum possible value and therefore the kinetic energy of superconducting electrons changes periodically with the magnetic field. It ought be noted that the using of the quantum force is not a principal new in comparison with the Tinkham's explanation. We may say: "the $T_{c}$ depends on $\Phi$ because the quantum force should be overcome at the superconducting transition if  $\Phi \neq n\Phi_{0}$". And we may say: "the $T_{c}$ depends on $\Phi$ because the energy of the superconducting state increases if $\Phi \neq n\Phi_{0}$". These statements are equivalent. Timkham did not consider the magnetic energy because $LI_{s}^{2}/2$ is proportional to $n_{s}^{2}$ and therefore does not influence on $T_{c}$.

The ring with narrow wall (the wall thickness $w \ll r, \lambda$, $\lambda$ is the penetration depth of magnetic field), when $LI_{s} \ll \Phi_{0}$ and $\Phi \simeq BS$, is considered in this paper. In the opposite case $w \gg \lambda$, $v_{s} \simeq 0$ and $\Phi = \int_{l} dl A \simeq n\Phi_{0}$ in the superconductor interior. $n$ is any integer number if a nonsuperconducting singularity is inside $l$ and $n = 0$ if a singularity is absent. The Meissner effect $\Phi  \simeq n\Phi_{0} = 0$ takes place in the later case. In order to describe this effect in the classical mechanic, the quantum force may be introduced also. But this description is not so obvious as in the case of the inhomogeneous superconducting ring. The Meissner effect is more intricate phenomenon than the quantization of the fluxoid: $n$ is not any integer number, but $n = 0$.

It is obvious that the direct voltage can appear only in the inhomogeneous case when the dissipating force $F_{dis}$ acts only in a segment of the ring. The quantum force $F_{q}$ accelerates electrons both in the $l_{a}$ and in the $l_{b}$ segments (only $\int_{t_{cl}}dt l^{-1}\int_{l}dl F_{q}$ has a sense), whereas the dissipating force $F_{dis}$ retards electrons only in the $l_{b}$ segment. Therefore the potential difference with dc component is induced in the ring segments. The $F_{q}$ acts only if the $F_{dis}$ acts. It returns the average momentum to the same value $<2p>_{l} = (2e/c)(\Phi_{0}/l)n$. The $<2p>_{l}$ value changes only at the switching of the segment $l_{b}$ into and out of the normal state. Any other changes of the superconducting electron density ($n_{sa}$ and $n_{sb}$) do not influence on this value.

\section{When and Why the Total Entropy may be Systematically Reduced}

The appearance of the direct voltage means that the inhomogeneous superconducting ring can be used as a direct-current generator. The current in a resistor $R_{load}$ loaded on the segment $l_{b}$ is $I_{load} = R_{b}I_{a}/(R_{b}+R_{load})$. After the transition of the segment $l_{b}$ in the normal state with $R_{b} = R_{bn}$ the current $I_{a}$ in the segment $l_{a}$ decreases exponentially from $I_{s}$ determined by the relation (3a) during the decay time $L/R_{sys}$, where $R_{sys} = R_{bn}R_{load}/(R_{bn} + R_{load})$. Consequently the power-driven energy $A_{sw} = \int dt R_{load}I^{2}_{load} = (R_{bn}/(R_{bn} + R_{load}))(LI_{s}^{2}/2)$ can develop across the load $R_{load}$ at the switching of the segment $l_{b}$. The power $W_{load} = A_{sw}f = (R_{bn}/(R_{bn} + R_{load}))F_{L}f$. Here $f = 1/N$ is the frequency of the switching; $N$ is the average number of the switching in a time unity.

Because the power $W_{load}$ may be utilized for a useful work and the segment $l_{b}$ may be switched by the temperature change, the superconducting ring may be used as the heat engine. The ordered force in this heat engine is the quantum force $F_{q}$. The $F_{q}$ direction coincides with the direction of the velocity $v_{s}$ of superconducting electrons. Whereas the dissipating force is directed against the velocity. The work done by the quantum force $\int dx F_{q} = \int dt v_{s}F_{q}$ increases the ordered energy: the kinetic energy of superconducting electrons and the energy of magnetic field $LI_{s}^{2}/2$. A part of this ordered energy may be used for a useful work and other part is dissipated in the ring after the $l_{b}$ switching into the normal state. The energy is dissipated completely if the load is absent i.e. $1/R_{load} = 0$.

There is not a contradiction with the second law of thermodynamics if the segment $l_{b}$ is switched in a consequence of a temperature change above and below $T_{cb}$ \cite{yalta98}. The heat energy $Q_{sw} = c_{b}\Delta T$ should be spent for the heating of the segment $l_{b}$ from $T_{min} = T_{cb} - \Delta T$ to $T_{max} = T_{cb}$. Here $c_{b}$ is the heat capacity. Because $I_{s} \propto n_{sb} \propto (T_{cb} - T)$, the work $A_{sw} \propto (T_{max} - T_{min})^{2}$. Consequently, the maximum efficiency $Ef = A_{sw}/Q_{sw}$ of the ring as a heat engine is proportional to $(T_{max} - T_{min})$. This is agreed with both the old and modern interpretation of the second law of thermodynamics. But the superconducting ring differs qualitatively from the conventional heat engine because it can work without correlation between $F_{ord}$ and $v$, i.e. it is not the type "a" heat engine, as the conventional heat engine, but is the type "b" heat engine. It can work at both a ordered and chaotic switching of the segment $l_{b}$.

The section $l_{b}$ can be switched chaotically by the thermal fluctuation at an unaltered temperature (at $T_{max} = T_{min} \simeq T_{cb}$). It is well known \cite{tink75,skocpo75} that the resistance of a superconductor $R < R_{n}$ in some region above $T_{c}$ and $R > 0$ in some region below $T_{c}$ because superconducting droplets (with characteristic size $\simeq \xi_{s}(T)$) appear in the normal state and normal droplets (phase-slip centers) (with characteristic size $\simeq \xi_{n}(T)$) appear in the superconducting state in the consequence of the thermal fluctuation. The coherence lengths $\xi_{s}(T) = \xi(0)(T/T_{c} - 1)^{-0.5}$ at $T > T_{c}$ and $\xi_{n}(T) = \xi(0)(1 - T/T_{c})^{-0.5}$ at $T < T_{c}$ in the linear approximation valid at $|T - T_{c}| \gg GiT_{c}$. $Gi$ is the Ginsburg number. We are interested here the one-dimensional case, in which the transverse dimensions of superconductor are small compared with the coherence length ($w < \xi$, $s < \xi^{2}$). In this case the $\xi_{s}(T)$ has a finite value at $T < T_{c}$, which increases with temperature decreasing. $\xi_{s}(T) \simeq \xi_{n}(T)$ at $T \simeq T_{c}$

The probability of the switching at $T \simeq T_{cb}$ of the segment $l_{b}$ in the normal state is much bigger than the one of the segment $l_{a}$ if $T_{ca} > T_{cb} \simeq T$. Therefore the inhomogeneous superconducting ring is switched by the fluctuation from closed ($n_{sa} > 0$, $n_{sb} > 0$) to open ($n_{sa} > 0$, $n_{sb} = 0$) superconducting state at an unaltered temperature corresponded to the resistive transition of the ring segment with the lowest critical temperature. As it was shown above, the voltage appears at this switching if $\Phi/\Phi_{0} \neq n$. The probability of the closing is enough high if $l_{b} \simeq \xi_{s}(T)$. The voltage is chaotic at $\Phi/\Phi_{0} = n + 0.5$, because the switching induced by the fluctuation is chaotic in time and the quantum force acts in opposite directions with equal probability. This case does not differ qualitatively from other fluctuation phenomena, for example from the Nyquist's noise \cite{Nyquist}. The power of the chaotic voltage is "spread" on all frequencies $\omega$ as well as at the Nyquist's noise, the power of which is proportional to a frequency band $\Delta \omega$: $<V_{Nyq}^{2}>/R = 4k_{B}T \Delta \omega$ \cite{kittel}.

The qualitative difference from the completely chaotic fluctuation effects (such as the Nyquist's noise) takes place at $\Phi/\Phi_{0} \neq n + 0.5$. n can be any integer number, but the state with minimum $|n - \Phi/\Phi_{0}|$ value has a maximum probability, because the energy of this state is minimum. Therefore the average value of the quantum force by a long time $T$ is not equal zero: $\int_{0}^{T}dt F_{q}/N = \sum_{sw.}P(|n - \Phi/\Phi_{0}|)(2e/c)(n\Phi_{0} - \Phi)/lN \neq 0$ at $\Phi/\Phi_{0} \neq n$ and $\Phi/\Phi_{0} \neq n + 0.5$. Here $N$ is the number of the switching during the time $T$; $\sum_{sw.}$ is the sum by these switching; $P(|n - \Phi/\Phi_{0}|)$ is the probability that $\int_{l} dl \bigtriangledown \phi/2\pi = n$ in the closed superconducting state. Because $\int_{0}^{T}dt F_{q}/N \neq 0$ the voltage $V_{b}$ is not completely chaotic ($V_{b}(\omega = 0) = <V_{b}> \neq 0$) although it is induced by the chaotic switching. This obvious consideration leads to result published first in \cite{jltp98}.

This result means that a part $p(\Phi/\Phi_{0})$ of the chaotic electric energy induced by thermal fluctuation can be ordered in the inhomogeneous superconducting ring. $p(\Phi/\Phi_{0}) = 0$ at $\Phi/\Phi_{0} = n$ or $\Phi/\Phi_{0} = n + 0.5$ and $0 < p(\Phi/\Phi_{0}) < 1$ at $\Phi/\Phi_{0} \neq n$ and $\Phi/\Phi_{0} \neq n + 0.5$. The direct voltage $V(\omega = 0) \neq 0$ can appear only in an inhomogeneous ring because in a homogeneous ring the switching is chaotic not only in time but also in space. Different segments of the ring are switched in different time (if $l \gg \xi(T)$). Therefore $<V> = 0$. In a mesoscopic ring with $l < \xi(T)$ the switching takes place simultaneously in the whole ring. Therefore the potential difference is equal zero.

The power of the energy regulating in the inhomogeneous superconducting ring is $W_{ord} = p(\Phi/\Phi_{0})A_{sw}f_{sw}$. $A_{sw} < F_{L}$ and the maximum frequency of the switching $f_{sw}$ is determined by the characteristic relaxation time of the superconducting fluctuation: $f_{sw} \leq 1/\tau_{rel.}$. Therefore  $W_{ord} < p(\Phi/\Phi_{0})F_{L}/\tau_{rel.}$. The fluctuations induce the magnetic energy $F_{L}
\simeq \mu k_{B}T/(1+\mu)$, where $k_{B}T$ is the characteristic energy of the thermal fluctuation; $\mu/(1+\mu)$ is the part of $F_{L}$ in
whole change of the superconductor energy. $\mu$ has the maximum value
$\mu = (32\pi^{3}/\kappa^{2})(Ls/l_{b}^{3})(n-\Phi/\Phi_{0})^{2}$ at
$T=T_{c}$. This relation is valid for a ring with $s \ll \lambda_{L}^{2}$.
Here $\kappa = \lambda_{L}/\xi$ is the superconductor parameter introduced
in the Ginsburg-Landau theory; $\lambda_{L}$ is the London penetration
depth of the magnetic field. In the linear approximation region at $T > T_{c}$, $\tau_{rel.} = \tau_{GL} = \hbar/8k_{B}(T-T_{c})$ (see \cite{skocpo75}). The probability of the
switching from into and out of the normal state is not
small only in the critical region, i.e. at $|T-T_{c}| < GiT_{c}$. Therefore
the maximum power of the energy regulating in the inhomogeneous superconducting ring may be estimated by the relation

$$W_{max} \simeq p(\Phi/\Phi_{0})\frac{\mu}{1+\mu} \frac{8\pi Gi (k_{B}T_{c})^2}{\hbar}
\eqno{(5)}$$
This power is very weak. Even for a high-Tc superconductor with $T_{c} \simeq 100 K$, $(k_{B}T_{c})^{2}/\hbar \simeq 10^{-8} Wt$.

Thus, the mesoscopic superconducting ring, switched by the fluctuation into and out of the closing superconducting state, may be considered as the second order perpetuum mobile of type "b". But the ring without a load are useless perpetuum mobile because the energy both is ordered and is dissipated inside it. Moreover, it is perpetuum mobile only in the old interpretation, which was revised in the beginning of our century together with the interpretation of the second law of thermodynamics. "Perpetuum mobile" is literally permanent motion. According to modern knowledge the permanent motion takes place at any nonzero temperature. The voltage inside the ring, as well as the Nyquist's noise, is one of the examples of this permanent motion.

This permanent motion is not perpetuum mobile according to the modern interpretation. It contradicts to old but not to modern interpretation of the second law of thermodynamics.
According to the old interpretation, the entropy decreases at the closing of the superconducting state and increase at the transition to normal state of the segment $l_{b}$. But according to the modern interpretation, the switching by the fluctuation does not change the entropy value, because both the closed and open superconducting state are included to the statistical weight $P$. The entropy, as the measure of chaos, does not change both at $<V_{b}> = 0$ and at $<V_{b}> \neq 0$ if whole magnetic energy $LI_{s}^{2}/2$ induced by the quantum force is dissipated in the segment $l_{b}$ after it's transition to the normal state, i.e. $<F_{q}><v_{s}> + <F_{dis}><v> = 0$. In this case, the situation in the inhomogeneous superconducting ring does not differ from the one in a homogeneous ring or at the Meissner effect.

Nevertheless the permanent motion in the mesoscopic superconducting ring differs qualitatively from other types of the permanent motion because it is partly ordered at $<F_{q}><v_{s}> \ > 0$. Therefore the potential chance exists to utilize this permanent motion for the useful work. In order to realize this chance the direct potential difference induced in the inhomogeneous superconducting ring by the quantum force should be put under load. The inhomogeneous superconducting ring with a load is the perpetuum mobile not only according to the old interpretation, but also the modern interpretation. It is the useful perpetuum mobile.
In this case no whole magnetic energy is dissipated in the segment $l_{b}$. A part develops across the load. This process contradicts to the Clausius's formula if a temperature of
the load is higher than the one of the ring. It contradicts to the
Carnot's principle and the Thomson's formula if the load is an
electricmotor. The total entropy (total chaos) is reduced when the heat energy is transformed in a ordered energy, $\Delta Q \rightarrow tW_{ord} \rightarrow E_{ord}$, because the reduction of the entropy $\Delta S = -\Delta Q/T = -E_{ord}/T$ can not be compensated in the equilibrium state for the transfer a heat energy from a heater to a cooler because $T_{heat} = T_{cool}$ in this state. The transformation $\Delta Q \rightarrow tW_{ord} \rightarrow E_{ord}$ can take place anyhow long time. At $tW_{ord} \gg k_{B}T$, $\Delta S \gg k_{B}$.

The useful work can be obtained in the load because the voltage induced by fluctuation is partly ordered, $V_{b}(\omega = 0) = <V_{b}> \neq 0$, in the inhomogeneous superconducting ring. There is qualitative difference, for example, from the Nyquist's noise \cite{Nyquist}, the power of which can not be used for a useful work because it is chaotic. It is impossible also to transfer any energy of the Nyquist's noise from a cold resistor to a hot resistor. The second law of thermodynamics can not be broken when the chaotic voltage is put under a load. The power of the Nyquist's noise is not summed up: the power of one resistor $<V_{Nyq}^{2}>/R = 4k_{B}T \Delta \omega$ is equal the one of N resistors $<V_{Nyq}^{2}>/NR = 4k_{B}T \Delta \omega$. Whereas the power $W_{ord}$ can be summed up: the direct voltage on a system of N rings, the segments of which with lowest critical temperature are connected in series, is equal $N<V_{b}>$. The power $W_{max}$ (see (5)) corresponds to the power of the Nyquist's noise in the frequency band $\Delta \omega = p(\Phi/\Phi_{0})(\mu/(1+\mu))2Gi(k_{B}T_{c}/\hbar)$. This $\Delta \omega$ may be enough wide. At $p(\Phi/\Phi_{0})(\mu/(1+\mu))2Gi \simeq 1$,  $\Delta \omega \simeq k_{B}T_{c}/\hbar \simeq 10^{13} \ sec^{-1}$ at $T_{c} = 100 K$.

The statement on the possibility of the perpetuum mobile is not new in essence. In the beginning of our century scientists have come to the conclusion that the perpetuum mobile, as the permanent motion of energy, takes place in any thermodynamic system at $T > 0$. The new statement in the my consideration is the possibility of regulating of this motion in some quantum system. This chaos reduction is connected with the quantum force, which has a fundamental cause: $\int_{l}dl p = \int_{l}dl \hbar \bigtriangledown \phi$ is "bad" (no gauge-invariant) quantum value if $l$ is not a closed path, and it is "good" (gauge-invariant) value if $l$ is a closed path. We may say that the quantum force acts at a "transition" from classical to quantum mechanics. It can be introduced because the dualism of electron. We can use the relation $mdv/dt = F$ because the superconducting electrons are particles. But these particles can accelerate against the classical forces because the electron is wave.

It is obvious that the entropy, as measure of the chaos, decreases at the "transition" from the classical to quantum mechanics. Planck has introduced in 1900 year the  quantization in order to reduce the number of microscopic states from infinite to finite value. Therefore one may say that the perpetuum mobile is possible because the "switching" between the classical and quantum mechanics is possible.

It was become obvious in the beginning of our century that the second law of thermodynamics in the old interpretation is correct to $k_{B}T$. The violation of the second order of thermodynamics in the modern interpretation has also the fundamental limit. Because the characteristic energy of fluctuation $k_{B}T$ and the time of any cycle can not be shorter than $\hbar/k_{B}T$ in accordance with the uncertainty relation, the power of any perpetuum mobile
$$W_{p.m.} < (k_{B}T)^{2}/\hbar \eqno{(6)}$$
We may say that the second law of thermodynamics is correct to $(k_{B}T)^{2}/\hbar $.

\section{How to Make the Perpetuum Mobile}
In order to make the perpetuum mobile, the modern methods of the nano-technology should be used. Sizes of the ring segment with lowest critical temperature should not surpass strongly the coherence length of the superconductor. Other sizes of the ring should be also enough small. Because the power of a ring is very weak, $W_{max} < (k_{B}T_{c})^{2}/\hbar$, a system with big number of the rings should be used in order to obtain an acceptable power. The useful power $W_{load} = N^{2} <V_{b}>^{2}R_{load}/(R_{load} + NR_{b})^{2}$ can be obtained by the system of N rings the segments $l_{b}$ of which are connected in series. This power has maximum value $W_{load} = N <V_{b}>^{2}/4R_{b}$ at $R_{load} = NR_{b}$. Because the sign and the value of $<V_{b}>$ depend on $n-\Phi/\Phi_{0} = n-BS/\Phi_{0}$, the area $S$ of all rings should be the same. The fluctuations induce $<V_{b}>$ only in a narrow region near $T_{cb}$. Therefore the critical temperature of all rings should be approximately identical. All rings should be identically inhomogeneous $T_{cb} < T_{ca}$. The requirement $T_{cb} < T_{ca}$ can be realized in the ring with difference areas of wall section $l_{a}$ and $l_{b}$, $s_{b} < s_{a}$. In accordance with the Little-Parks effect $T_{cb}(\Phi) < T_{ca}(\Phi)$ at $s_{b} < s_{a}$ if $\Phi/\Phi_{0} \neq n$.

Because $W_{max} \propto T_{c}^{2}$ a system of high-Tc superconductor (HTSC) rings has maximum power. In order to obtain $W_{load} \approx 1 \ Wt$ the system of no less than $4 \ 10^{8}$ HTSC rings should be used. But it is very difficult to make the HTSC rings with enough small sizes. The coherence length of all known HTSC is very small. Therefore the conventional superconductor rings (for example Al rings) ought be used for first experimental investigation. The expected voltage $<V_{b}>  \ < \ (R_{b}W_{max})^{1/2} \simeq (p(\Phi/\Phi_{0})8 Gi\mu/(1+\mu))^{1/2} R_{b}^{1/2} k_{B}T_{c}/\hbar^{1/2}$ is small but measurable. For example at $T_{c} = 1 \ K$ and a real value $R_{b} = 1 \ \Omega$, $R_{b}^{1/2} k_{B}T_{c}/\hbar^{1/2} \simeq 10^{-6} \ V = 1 \ \mu V$. $(p(\Phi/\Phi_{0})8 Gi\mu/(1+\mu))^{1/2} \leq 1$ in any case. This value depends in the periodic manner on the magnetic field (with period $\Delta B = \Phi_{0}/S$) and is not equal zero only at the temperature closed to $T_{cb}$.

The perpetuum mobile can not solve the energy problem. It's power is very weak. But the system of superconducting rings may be used in some applications. It can used simultaneously as the direct-current generator \cite{Turkey99} and as the micro-refrigerator \cite{Jyvaskyla}. The perpetuum mobile can work anyhow long time without an expense of any fuel. Therefore it may be especially useful in self-contained systems.

\section{ACKNOWLEDGMENTS} I am grateful Jorge Berger for the preprint of
his paper and for stimulant discussion. I thank for financial support the
International Association for the Promotion of Co-operation with Scientists
from the New Independent States (Project INTAS-96-0452).

\

\end{document}